\documentclass[journal]{IEEEtran}
\usepackage{graphicx}
\usepackage{amsmath}
\usepackage{amsfonts}
\usepackage{breqn}
\usepackage{multirow}
\usepackage{tikz}
\usepackage{subfig}
\usetikzlibrary{patterns}
\usetikzlibrary{arrows}
\usetikzlibrary{shapes,snakes}
\usetikzlibrary{backgrounds,fit,decorations.pathreplacing}
\usepackage{graphicx}
\usepackage{caption}
\usepackage{pgfplots}
\pgfplotsset{grid style={dashed, gray}}
\usetikzlibrary{shapes,arrows,patterns}

\ifCLASSINFOpdf
\else
\fi

\begin{document}
\bstctlcite{IEEEexample:BSTcontrol}
\title{Bounds for Energy Efficient Survivable IP over WDM Networks with Network Coding}

\author{Mohamed~Musa,
        Taisir~Elgorashi,
        and~Jaafar~Elmirghani}% <-this % stops a space
%\thanks{Manuscript received April 19, 2005; revised January 11, 2007.}}

%\markboth{Journal of \LaTeX\ Class Files,~Vol.~6, No.~1, January~2007}%
%{Shell \MakeLowercase{\textit{et al.}}: Bare Demo of IEEEtran.cls for Journals}

\maketitle

\begin{abstract}
%\boldmath
In this work we establish analytic bounds for the energy efficiency of the 1+1 survivable IP over WDM networks using network coding. The analytic bounds are shown to be in close agreement with our previously reported results. They provide verification of the MILP and heuristics proposed previously and an efficient, compact means of evaluating the network results, and allow the performance of large networks to be determined easily.  
\end{abstract}


\IEEEpeerreviewmaketitle

\section{Introduction}

\IEEEPARstart{A}{fter} the introduction of network coding for the first time in \cite{Ahlswede2000}, the contributions of network coding (NC) to various networking domains have accelerated, demonstrating the potential it has to improve the networking throughput. The work in optical networks, however, has been incomparable to its wireless counterpart due to the multicast nature of the wireless medium that is not inherent in the optical medium. 
%This is the reason why most of the work in network coding in optical networks focused on the passive optical networks and protection. 
The ability of network coding to reduce the overall traffic in the network, and therefore improve the network throughput, provides a motivation for using network coding to achieve energy efficiency by requiring less operating resources than the conventional approach. The benefits of introducing network coding (NC) in optical networks to improve robustness and efficiency has been reported in  \cite{kamal2014network}, \cite{babarczi2013realization}, and  \cite{overby2012cost}. In our previous work \cite{musa2015energy} and \cite{musa2016energy} we studied the energy efficiency gained by implementing network coding in non-bypass and bypass core networks by performing an XOR (Exclusive OR) operation on the bidirectional flows of unicast connections.  Network coding elevates the traditional functionality of network nodes to incorporate coding operations on traffic flows and hence, by mixing signals at specific nodes rather than duplicating the signals end to end, more efficient network resource utilization can be achieved.\par
In \cite{kamal2011efficient} the authors provided a $1+N$ network coding protection scheme, and through integer linear programmes and simulation they showed that significant cost savings over the $1+1$ approach can be achieved.
Network coding was proposed in \cite{kamal2010network} and \cite{aly2008network} as a technique to improve protection in $1+N$ protection schemes that employ p-cycles. The p-cycles are used to protect multiple bidirectional link-disjoint connections, which are also link disjoint from the p-cycle links.
In \cite{aly2009network}, network coding is used to provide protection against node failures by reducing the problem to a problem of multiple link failures as a consequence of the node failure. In \cite{barla2010network} it is shown that for networks with multiple subdomains, network coding can be used to enable the network to survive any node or link failure in each subdomain.
The study of $1+1$ protection schemes with network coding was reported in \cite{muktadir2012optimum}, through an integer non linear programme. This study is limited however, to equal traffic demands between different sources, provides results that are considerably lower than those achievable through network coding, and constrains the network coding only to nodes with nodal degree greater than or equals to 3. Our work is different in that it focuses on the widely implemented 1+1 protection scheme where it provides optimal and thorough solutions to protection with network coding focusing on improving the energy efficiency of the network. As far as our knowledge goes, no practical implementation is found yet in the industry, but we are optimistic with the considerable resource savings that network coding can achieve, the implementation will follow.  \par
% Moreover, much of research on NC-based protection for optical network is limited to opaque (O-E-O) architecture where the NC could be performed in electrical domain together with electronic buffering and processing at each node \cite{kamal2014network}, \cite{babarczi2013realization} and \cite{overby2012cost}. This leaves a gap in literature on whether NC could be useful in transparent/all-optical network and with  the maturing of all-optical processing and the emerging of transparent architecture makes this question particularly pertinent. \par
Energy efficiency in IP over WDM network has attracted a considerable attention from the research community driven by economic and environmental impact. The exponential growth of data intensive applications and the increasing number of Internet connected devices necessitates a shift in the way the network is designed and operated. As a global effort to tackle the energy consumption challenge in ICT, the GreenTouch consortium of leading expert in Industry and academia was formed in 2010 with the goal of achieving a 1000x energy efficiency improvement in 2020 compared to 2010 levels. The GreenTouch results for the core network are reported in \cite{Greentouch2017}. A good survey of some of the techniques for energy efficiency in core networks can be found in  \cite{idzikowski2016survey,dharmaweera2015toward}. We used MILP models and heuristics in our previous work to improve energy efficiency in IP over WDM networks considering renewable energy sources \cite{dong2011ip},  studying core networks with data centers \cite{Dong2011}, optimising physical topology \cite{Dong2012}, reducing traffic through distributed clouds \cite{LaweyCloud}, optimum design for future high definition TV \cite{Niema2014}, optimal P2P content distribution \cite{Lawey2014} and virtual network embedding \cite{Nonde2015}. We introduced network coding for energy efficient IP over WDM networks in  \cite{musa2015energy} and \cite{musa2016energy}, by encoding bidirectional flows using an XOR operation, and presented a thorough study of  the use of network coding to improve the energy efficiency in core networks in unicast settings \cite{MusaJLT}.\par
In our previous work \cite{musa2017energy} we proposed and designed a 1+1 protection scheme for core networks with network coding and optimized the network coding allocation and network operation using a mixed integer linear program and heuristics, providing encouraging results for energy efficiency improvement of up to 37\% compared to the conventional 1+1 protection scheme. In this work we complement our study and analysis by deriving for the first time analytical bounds and close form expressions for the network coding as well as the conventional 1+1 protection scheme which verify the MILP and heuristic results in \cite{musa2017energy} and enable the performance of large networks to be determined easily. We also study large network sizes that are highly complex using the MILP approach, as well as provide a detailed study on the special full mesh and ring topologies

%In our work \cite{musa2017energy} we studied 1+1 protection with network coding in core networks through a mixed linear integer program and heuristics, providing encouraging results for energy efficiency improvement of up to 37\% compared to the conventional 1+1 protection scheme. In this paper we derive for the first time analytical bounds and close form expressions for our network coding scheme which verify the MILP and heuristic results in \cite{musa2017energy} and enable the performance of large networks to be determined easily. We study large network sizes, as well as provide a detailed study on the special full mesh and ring topologies. \par
The remainder of this paper is organized as follows: Section II provides the analytical bounds for the conventional and network coded core networks with 1+1 protection.  In Section III we derive the bounds for regular topologies. Finally the paper is concluded in Section IV.
\section{1 + 1 PROTECTION WITH NETWORK CODING}
\begin{figure}[h]
\centering
    \begin{tabular}{cccc}
    \subfloat[]{
        \begin{tikzpicture}[scale=0.55]
         \tikzstyle{obj}  = [circle, minimum width=5pt, draw,fill=white, inner sep=0.1cm, font=\scriptsize]
          \tikzstyle{obj_scr}  = [circle, minimum width=5pt, draw,fill=red!20, inner sep=0.1cm, font=\scriptsize]
         %\tikzstyle{every node}=[draw,circle,fill=white,minimum size=6pt,
                                   % inner sep=0.2cm]
        \tikzstyle{obj2}=[ midway, above, fill=white, font=\scriptsize]
            \draw (0,0) node[obj_scr] (2)  {2}
                 ++(0:3.0cm) node[obj] (1)  {1}
                 ++(0:3.0cm) node[obj_scr] (3) {3}
                 ++(270:3.0cm) node[obj] (6)  {6}
                 ++(270:3.0cm) node[obj] (7)  {7}
                 ++(225:4.25cm) node[obj_scr] (11)  {11}
                 ++(90:2.5cm) node[obj] (10) {10}
                 ++(90:2.0cm) node[obj] (9)  {9}
                 ++(90:2.0cm) node[obj] (8)  {8}
                 (1)
                 (2)
                 ++(270:3.0cm) node[obj] (4)  {4}
                 ++(270:3.0cm) node[obj] (5)  {5};
        \draw (2) --(1); 
        \draw (2) --(4);
        \draw (1) --(3);
        \draw (3) --(6);
        \draw (6) --(7);
        \draw (7) --(11);
        \draw (4) --(5);
        \draw (5) --(11);
        \draw (1) --(8);
        \draw (8) --(9);
        \draw (9) --(10);
        \draw (10) --(11);
        \draw[->,red] (2.20) --(1.160) node[obj2] {$\lambda1$};
        \draw[->,blue] (3.160) --(1.20) node[obj2] {$\lambda2$}; 
        \draw[->,red] (3.290) --(6.70) node[obj2] {$\lambda1$}; 
        \draw[->,red] (6.290) --(7.70) node[obj2] {$\lambda1$}; 
        \draw[->,red] (7.250) --(11.20) node[obj2,below] {$\lambda1$}; 
        \draw[->,red] (2.250) --(4.110) node[obj2] {$\lambda1$}; 
        \draw[->,red] (4.250) --(5.110) node[obj2] {$\lambda1$}; 
        \draw[->,red] (5.290) --(11.160) node[obj2,below] {$\lambda1$};
        \draw[->,blue] (1.290) --(8.70) node[obj2,right] {$\lambda2$}; 
        \draw[->,red] (1.250) --(8.110) node[obj2] {$\lambda1$};
        
        \draw[->,blue] (8.290) --(9.70) node[obj2,right] {$\lambda2$}; 
        \draw[->,red] (8.250) --(9.110) node[obj2] {$\lambda1$}; 
        \draw[->,blue] (9.290) --(10.70) node[obj2,right] {$\lambda2$}; 
        \draw[->,red] (9.250) --(10.110) node[obj2] {$\lambda1$}; 
        \draw[->,blue] (10.290) --(11.70) node[obj2,right] {$\lambda2$}; 
        \draw[->,red] (10.250) --(11.110) node[obj2] {$\lambda1$};
       \end{tikzpicture}}&
\subfloat[]{
        \begin{tikzpicture}[scale=0.55]
         \tikzstyle{obj}  = [circle, minimum width=5pt, draw,fill=white, inner sep=0.1cm, font=\scriptsize]
         \tikzstyle{obj3}  = [circle, minimum width=5pt, draw,fill=green, inner sep=0.1cm, font=\scriptsize]
          \tikzstyle{obj_scr}  = [circle, minimum width=5pt, draw,fill=red!20, inner sep=0.1cm, font=\scriptsize]
         %\tikzstyle{every node}=[draw,circle,fill=white,minimum size=6pt,
                                   % inner sep=0.2cm]
        \tikzstyle{obj2}=[ midway, above, fill=white, font=\scriptsize]
            \draw (0,0) node[obj_scr] (2)  {2}
                 ++(0:3.0cm) node[obj3] (1)  {1}
                 ++(0:3.0cm) node[obj_scr] (3) {3}
                 ++(270:3.0cm) node[obj] (6)  {6}
                 ++(270:3.0cm) node[obj] (7)  {7}
                 ++(225:4.25cm) node[obj3] (11)  {11}
                 ++(90:2.5cm) node[obj] (10) {10}
                 ++(90:2.0cm) node[obj] (9)  {9}
                 ++(90:2.0cm) node[obj] (8)  {8}
                 (1)
                 (2)
                 ++(270:3.0cm) node[obj] (4)  {4}
                 ++(270:3.0cm) node[obj] (5)  {5};
        \draw (2) --(1); 
        \draw (2) --(4);
        \draw (1) --(3);
        \draw (3) --(6);
        \draw (6) --(7);
        \draw (7) --(11);
        \draw (4) --(5);
        \draw (5) --(11);
        \draw (1) --(8);
        \draw (8) --(9);
        \draw (9) --(10);
        \draw (10) --(11);
        \draw[->,red] (2.20) --(1.160) node[obj2] {$\lambda1$};
        \draw[->,red] (3.160) --(1.20) node[obj2] {$\lambda1$}; 
        \draw[->,red] (3.290) --(6.70) node[obj2] {$\lambda1$}; 
        \draw[->,red] (6.290) --(7.70) node[obj2] {$\lambda1$}; 
        \draw[->,red] (7.250) --(11.20) node[obj2,below] {$\lambda1$}; 
        \draw[->,red] (2.250) --(4.110) node[obj2] {$\lambda1$}; 
        \draw[->,red] (4.250) --(5.110) node[obj2] {$\lambda1$}; 
        \draw[->,red] (5.290) --(11.160) node[obj2,below] {$\lambda1$};
        \draw[->,red] (1.290) --(8.70) node[obj2] {$\lambda1$}; 
        \draw[->,red] (8.290) --(9.70) node[obj2] {$\lambda1$}; 
        \draw[->,red] (9.290) --(10.70) node[obj2] {$\lambda1$}; 
        \draw[->,red] (10.290) --(11.70) node[obj2] {$\lambda1$}; 
        \end{tikzpicture}}&    
        \end{tabular}
    \caption{Example of using network coding for protection}
    \label{fig:NC11}
    \end{figure}
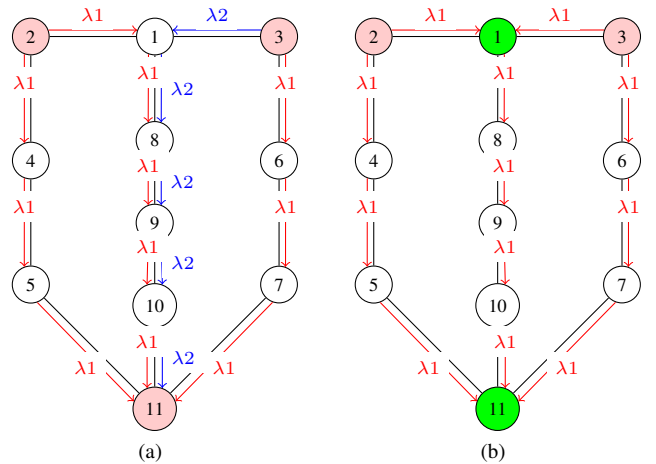

Consider the network coding scheme where an example is shown in Fig. 1, representing a comparison between the conventional (Fig.
\ref{fig:NC11}.a) and the network coded (Fig. \ref{fig:NC11}.b) 1+1 protection scheme in an arbitrary topology. Consider the two demands (3, 11) and (2, 11) originating from node 3 and 2 respectively and sharing the same destination. With the conventional 1+1 protection scheme (Fig. \ref{fig:NC11}.a) both demands use a single wavelength ($\lambda_1$) for the working path and each working path traverses three links. Since they share the 4 links of their protection paths, they are forced to use different wavelengths, ($\lambda_1$) for the demand (2,11) and ($\lambda_2$) for demand (3,11), using a total of 16 wavelengths in the network, considering all links and paths. \\
Our proposed network coding scheme is shown in (Fig. \ref{fig:NC11}.b), where the protection paths use the same wavelength ($\lambda_1$) after being encoded at node 1, and later decoded at the destination node (i.e. 11). The benefits of using such a scheme are in reducing the total number of distinct wavelength used in the network, where in this case only the wavelength ($\lambda_1$) is used, as well as reducing the total number of wavelengths in the network; in this example from 16 to 12. A reduction of 25\% in total resources, and 40\% in protection resources (6 wavelengths instead of 10). \\
The resource savings and therefore the power consumption reduction in this scheme depends on the network topology, the location and number of network coding enabled nodes as well as the nature of demands. In our previous paper [23] we studied the scheme and determined the optimum allocation using a MILP optimization model followed by 5 heuristics. 
 The real time and most optimal version of the heuristics is called the Optimal search heuristic (OSH). The heuristic reduces the size of the problem by dividing the set of encodable paths into clusters and searching for optimal coding operations on these clusters rather than the whole network. The working and protection paths are found using the Suurbelle algorithm. Four other suboptimal but faster heuristics are found by limiting the search for a suitable encoding pair. For each pair of demands there exists 4 paths in total; a working path and protection path for each. This produces 4 combinations for encoding, and therefore 4 possible specific heuristics, we call them w-w, p-p, w-p and p-w where the letters w and p designate the working and protection path respectively. The heuristics assume a distributed approach to determine the encoding decision. The network state can be communicated using the conventional routing protocol mechanism to exchange network state. For higher order codes and coding more than two paths, a centralized control may have more value.  It would be interesting to consider the SDN and virtualization ideas presented in \cite{Akhilesh2016} and \cite{capone2015detour} for optical networks generally and for protection and failure recovery specifically as a future area of investigation.\\
In this work we complement our work by developing closed form expressions and analytic bounds for the power consumptions as a function of the hop count, the network size and the demand volume. We also study regular topologies and study scenarios that proved too complex for the MILP approach such as large network sizes. 
%Considering the scheme presented in the previous paper \cite{musa2017energy}, where an example is shown in Fig.  \ref{fig:NC11}, representing a comparison between the conventional (Fig.
%\ref{fig:NC11}.a) and the network coded (Fig. \ref{fig:NC11}.b) 1+1 protection scheme in
%an arbitrary topology. For the two demands $(3,11)$ and $(2,11)$, a reduction in the total number of used wavelengths in the network from 16 wavelengths down to 12 wavelengths by encoding the protection paths of both flows together using an xor operation at node $1$ and decoding it at the destination. 
\section{Analytic bounds}
 The power consumption of the network is calculated as the sum of the power consumption of individual components. This approach is used by the research community and adopted by the Greentouch consortium \cite{Greentouch2017}. 
To simplify the analytical formulas, we consider the network devices components to have a linear power profile where the power consumption of a device is proportional to the traffic served. Other power profiles exist and have been investigated in our previous work \cite{dong2011ip}, such as the on-off, cubic and log10 profiles.\\
The total power consumption of the survivable optical networks with network coding for this scheme is given by
\begin{multline}
  P=(\frac{p_p+p_t}{B})  \sum_{d\in D}\sum_{m,n}V^{d}(x_{mn}^{d}+y_{mn}^{d})\\- (\frac{p_p+p_t}{B})\sum_{d1,d2}\sum_{m,n}\min(V^{d_1},V^{d_2})\frac{\beta_{mn}^{d_1d_2}}{2}, 
  \label{eq:totPower}  
\end{multline}
where the first term is the total power consumption of the network operating without network coding, and the second term is the power consumption reduction achieved by network coding. $p_p, p_t$ are the power consumption of a router port and a transponder in Watts, $B$ is the capacity of a wavelength in Gbps, $V^d$ is the volume of demand $d$ in Gbps, $D$ is the set of demands. The variable $x^d_{m,n}$ is a Binary variable, $x^{d}_{mn}=1$ if the working path of demand $d$ is routed over link $(m,n)$, and $x^{d}_{mn}=0$ otherwise. The variable $y^d_{m,n}$ is the equivalent of $x^d_{m,n}$ for protection paths. $\beta^{d_1d_2}_{mn} $  is a binary variable, $\beta^{d_1d_2}_{mn}=1$ if demand $d_1$ is encoded with demand $d_2$ on link $(m,n)$, and $\beta^{d_1d_2}_{mn}=0$ otherwise. Note that the  values of power consumption ($p_p$ and   $p_t$) count for the OEO conversion at all nodes. The routing from source to destination uses the optical non-bypass approach where all intermediate nodes have OEO conversion.
The power consumption contributions of the XOR operations, and of the EDFAs have not been included as their associated power consumption is low and to simplify the expressions. According to the Greentouch core network energy efficiency study \cite{Greentouch2017},  the EDFAs power consumption constitutes a small portion of the overall power consumption compared to the routers and transponders power consumption. For the 2010 values the EDFAs consume 5\% of the overall power consumption, while for the predicted business as usual values for 2020 they consume less than 2\% at port speeds of 400Gbps\par
Let the expression given in (\ref{eq:totPower}) be divided
into its two summable components, which we refer to  as $P_{1}$ and
$P_{2}$ (i.e. $P=P_{1}-P_{2}$), such that
\begin{equation}
P_{1}=(\frac{p_p+p_t}{B})\left(\sum_{d\in D}\sum_{m,n}V^{d}(x_{mn}^{d}+y_{mn}^{d})\right)\label{eq:sumXY},
\end{equation}
\begin{equation}
P_{2}=(\frac{p_p+p_t}{B})\left(\sum_{d1,d2}\sum_{m,n}\min(V^{d_1},V^{d_2})\frac{\beta_{mn}^{d_1 d_2}}{2}\right).
\label{eq:Prot}
\end{equation}
The value $P_{1}$ represents the power consumption of the baseline conventional 1+1 protection approach, while $P_{2}$ is the reduction as a result of using network coding. \\
Expression (\ref{eq:sumXY}) can be rewritten as
\begin{equation}
P_{1}=(\frac{p_p+p_t}{B})\left(\sum_{d\in D}V^{d}\sum_{m,n}(x_{mn}^{d}+y_{mn}^{d})\right).
\label{eq:sumXY2}
\end{equation}
Given the fact that the sum of the hop count of the working and protection paths of a given demand is always greater than or equal to twice the minimum hop count $h^{d}_{min}$ of the path serving it, that is
\begin{equation}
\sum_{m,n}(x_{mn}^{d}+y_{mn}^{d})\ge 2 h^{d}_{min},
\label{eq:XYHop}
\end{equation}
therefore $P_{1}$ can be written as
\begin{equation}
P_{1}\ge(\frac{p_p+p_t}{B})\left(\sum_{d\in D}2V^{d}h^{d}_{min}\right).
\label{eq:sumXY3}
\end{equation}
Assuming that all demands are routed through the minimum hop count path of the network, i.e. $h^{d}_{min}=h_{min}, \forall d \in D$, we then  have
\begin{equation}
P_{1}\ge(\frac{p_p+p_t}{B})h_{min}\left(\sum_{d\in D}2V^{d}\right),  
\label{eq:sumXY3_2}
\end{equation}
which gives
\begin{equation}
P_{1}\ge 2(\frac{p_p+p_t}{B})N(N-1)V h_{min}. 
\label{eq:sumXY3_3}
\end{equation}
Expression (\ref{eq:sumXY3_3}) represents a lower bound on the power consumption of the first component of the total power consumption of the network coded case as a result of routing traffic flows in the working and the protection paths, without the network coding component. It also represents the lower bound on the power consumption of the conventional case, which we refer to as $P_{0}$, where
%Referring to the power consumption of the conventional case as $P_0$, it is found that it has the following lower bound, where $\langle \textbf{V}, \textbf{h} \rangle $ represents the inner product of the demand values vector and the demands minimum hop count vector.  \par
\begin{equation}
P_{0}\ge 2(\frac{p_p+p_t}{B})N(N-1)V h_{min}. 
\label{eq:sumXY4}
\end{equation}
The upper bound of $P_2$ is found by starting from the fact that the minimum volume of two demands is never greater than their average, that is
\begin{equation}
\min(V^{d_1},V^{d_2})\le \frac{V^{d_1}+V^{d_2}}{2}, 
\label{eq:minV1V2}
\end{equation}
then (\ref{eq:Prot}) becomes
\begin{equation}
P_{2}  \le (\frac{p_p+p_t}{B})\left(\sum_{d1,d2}\sum_{m,n}\frac{V_{d_1}+V_{d_2}}{2}\frac{\beta_{mn}^{d_1 d_2}}{2}\right). \label{eq:Prot-1}
\end{equation}
The expression $\min(V^{d_1},V^{d_2})$ has its highest value when the maximum traffic is equal to the minimum traffic, therefore the equality in (\ref{eq:minV1V2}) is met when $V^{d_1}=V^{d_2}=V^{d_1,d_2}$. In this case (\ref{eq:Prot-1}) becomes
\begin{equation}
P_{2}\le \frac{p_p+p_t}{2B}\left(\sum_{d_1,d_2}\sum_{m,n}V^{d_1,d_2}\beta_{mn}^{d_1 d_2}\right).
\label{eq:Prot-1}
\end{equation}
This can be rearranged as
\begin{equation}
P_{2}\le \frac{p_p+p_t}{2B}\left(\sum_{d_1,d_2}V^{d_1,d_2}\sum_{m,n}\beta_{mn}^{d_1 d_2}\right).
\label{eq:Prot-2}
\end{equation}
The expression $\sum_{m,n}\beta_{mn}^{d_1 d_2}$ represents the number of shared links between the demand pair ($d_1,d_2$). We refer to this value as $h_{d_2}^{d_1}$, where 
\begin{equation}
h_{d_2}^{d_1}=\sum_{m,n}\beta_{mn}^{d_1 d_2}\label{eq:alpha}.
\end{equation}
Therefore equation (\ref{eq:Prot-2}) becomes:
\begin{equation}
P_{2}\le \frac{p_p+p_t}{2B}\left(\sum_{d_1,d_2}V^{d_1,d_2}h_{d_2}^{d_1}\right).
\label{eq:Prot-3}
\end{equation}
Considering the lower bound of the component $P_1$ in (\ref{eq:sumXY3}) and the upper bound of the component $P_2$ in (\ref{eq:Prot-3}), we can get a lower bound on the total power consumption by combining the two, since $P=P_1-P_2$, minimising $P$ is achieved by minimising $P_1$ and maximising $P_2$. The total power is then lower bounded by the following
\begin{equation}
P\ge(\frac{p_p+p_t}{B})\left(2 \sum_{d \in D}V^d h^d_{min} - \frac{1}{2}\sum_{d_1,d_2}V^{d_1,d_2}h_{d_2}^{d_1}\right).
\label{eq:totPower2}
\end{equation}
The result in (\ref{eq:totPower2}) analytically confirms our heuristic in \cite{musa2017energy} that can provide close to optimal solution. The heuristic produces a good solution by employing the following principles 
\begin{itemize}
    \item Select the minimum number of hops for the working and protection paths (minimising the first term of (\ref{eq:totPower2}))
    \item Encode a demand with another demand that has the highest number of shared hops and closest demand volume (maximising the second term of (\ref{eq:totPower2}))
    \item More weight is given to finding minimal hop paths than searching for better encoding pair (from the equation, the weight ratio of the first to second terms is 4:1)
    \item Three heuristics can be conceived. The first finds encodable pairs by searching only for the highest link sharing, the second searches for the demand with the closest traffic volume, and a better heuristic searches for the highest sharing and closest traffic volume, at the expense of increased complexity. The first heuristic approaches the performance of the third the smaller the traffic variation becomes. 
\end{itemize}
%which can be written as:
%\begin{equation}
%P\ge(\frac{p_p+p_t}{B})\sum_{d\in D}\left( 2V_{d}h_{d}-\frac{1}{2}\sum_{d2}h_{d2}^{d}V_{d,d2}\right)\label{eq:totPower3}
%\end{equation}
The bound (\ref{eq:totPower2}) can be reduced by setting $V^{d}=V^{d,d_2}$, which gives
\begin{equation}
P\ge(\frac{p_p+p_t}{B})\sum_{d\in D}V^{d}\left(2h^d_{min}-\frac{1}{2}\sum_{d_2}h_{d_2}^{d}\right).
\label{eq:totPower4}
\end{equation}
Since each demand is constrained to be encoded with a maximum of a
single other demand only, which is expressed as
\begin{equation}
\sum_{d2\in D} b_{d_2}^{d_1}\le1\label{eq:sumBeta}.
\end{equation}
Therefore we let the value $\hat{h}^d=\sum_{d_2}h_{d_2}^{d}$ represent the amount of shared links (hops) between demand $d$ and the demand it is encoded with. 
Equation (\ref{eq:totPower4}) can then be reduced to
\begin{equation}
P\ge(\frac{p_p+p_t}{B})\left(\sum_{d\in D}V^{d}(2h^{d}_{min}-\frac{\hat{h}^d}{2})\right), 
\label{eq:totPower5}
\end{equation}
which is equal to 
\begin{equation}
P\ge2(\frac{p_p+p_t}{B})\left(\sum_{d\in D}V^{d}(h^{d}_{min}-\frac{\hat{h}^d}{4})\right).
\label{eq:totPower6}
\end{equation}
If we define the variable $\tilde{h}^d$ as the characteristic hop count for demand $d$, such that
\begin{equation}
\tilde{h}^d=h^{d}_{min}-\frac{\hat{h}^d}{4}, 
\label{eq:Gamma}
\end{equation}
therefore the lower bound of the total power becomes
\begin{equation}
P\ge2(\frac{p_p+p_t}{B})\left(\sum_{d\in D}V^{d}\tilde{h}^d\right).
\label{eq:totPower7}
\end{equation}
Using Chebyshev Sum Inequality, i.e.
\begin{equation}
    \frac{1}{n} \sum_{k=1}^{n} a_k b_k \ge \left( \frac{1}{n}\sum_{k=1}^{n} a_k \right)\left( \frac{1}{n}\sum_{k=1}^{n} b_k \right), 
\end{equation}
then, (\ref{eq:totPower7}) can be written as
\begin{equation}
P\ge2(\frac{p_p+p_t}{B})\left( \frac{1}{N(N-1)}\sum_{d\in D}V^{d}\sum_{d \in D}\tilde{h}^d\right), 
\label{eq:totPower7_2}
\end{equation}
which gives
\begin{equation}
P\ge2(\frac{p_p+p_t}{B})V\sum_{d\in D}\tilde{h}^d,
\label{eq:totPower10}
\end{equation}
where $V=\sum_{d \in D} V^d$ is the average demand volume. Defining $\tilde{h}=\sum_{d\in D}\tilde{h}^d$ as the average characteristic hop count, then 
\begin{equation}
P\ge2(\frac{p_p+p_t}{B})N(N-1)V\tilde{h}\label{eq:totPower10}
\end{equation}
The lower bound given in (\ref{eq:totPower10}) bears resemblance to the lower bound of the conventional case in (\ref{eq:sumXY4}), where the minimum hop count of the conventional case $h_{min}$ is replaced by the characteristic minimum hop count of the network coding case $\tilde{h}$. 

\section{Regular topologies}
In the previous work \cite{musa2017energy} we established that the star and the line topologies exhibits no network coding benefits as the concept of protection does not apply. Here we develop formulas and bounds for the full mesh and ring topologies for the case where protection paths are encoded together, and study the impact of the network size on the performance of network coding. 
\subsection{Full mesh topology}

The total power consumption under conventional protection is given by
\begin{equation}
P_0=(\frac{p_p+p_t}{B})\left(\sum_{d\in D}\sum_{m,n}V^{d}(x_{mn}^{d}+y_{mn}^{d})\right)\label{eq:fullmesh1}
\end{equation}
For the full mesh topology, the optimal paths are the direct path (a single hop) for the working path, and a path with an intermediate node for the protection path ($2$ hops). This means ($\sum_{m,n}x_{mn}^d=1$) and ($\sum_{m,n}y_{mn}^d=2$), $\forall d \in D$. Therefore

\begin{equation}
P_0=(\frac{p_p+p_t}{B}) \left(\sum_{d\in D}3 V^d\right),
\label{eq:fullmesh2}
\end{equation}
which can be written as 
\begin{equation}
P_0=3(\frac{p_p+p_t}{B})V N(N-1).
\label{eq:fullmesh3}
\end{equation}

\begin{figure*}[ht]
\begin{minipage}[b]{0.3\linewidth}
\centering
\begin{tikzpicture}[scale=1]
            \tikzstyle{obj}  = [regular polygon,regular polygon sides=8, minimum width=10pt, draw,fill=white, inner sep=0.1cm, font=\scriptsize]
            \tikzstyle{obj1}  = [regular polygon,regular polygon sides=8, minimum width=10pt, draw,fill=green!20, inner sep=0.1cm, font=\scriptsize]
            \tikzstyle{obj2}=[ midway, above, fill=white, font=\scriptsize]
            \draw (0,0) node[obj1] (1)  {1}
                 ++(0:2.0cm) node[obj1] (2)  {2}
                 ++(288:2.0cm) node[obj] (3) {3};
            
            \path[<->] (1) edge node[obj2] {1} (2);

        \end{tikzpicture}
\label{fig:figure1}
\end{minipage}
\begin{minipage}[b]{0.3\linewidth}
\centering
\begin{tikzpicture}
            \tikzstyle{obj}  = [regular polygon,regular polygon sides=8, minimum width=10pt, draw,fill=white, inner sep=0.1cm, font=\scriptsize]
            \tikzstyle{obj1}  = [regular polygon,regular polygon sides=8, minimum width=10pt, draw,fill=green!20, inner sep=0.1cm, font=\scriptsize]
            \tikzstyle{obj2}=[ midway, above, fill=white, font=\scriptsize]
            \draw (0,0) node[obj1] (1)  {1}
                 ++(0:2.0cm) node[obj1] (2)  {2}
                 ++(288:2.0cm) node[obj1] (3) {3}
                 ++(216:2.0cm) node[obj1] (4)  {4};
                 
                \path[<->] (1) edge node[obj2] {1} (2);
                \path[<->] (3) edge node[obj2] {1} (4);
                
        \end{tikzpicture}
\label{fig:figure2}
\end{minipage}
\begin{minipage}[b]{0.3\linewidth}
\centering
\begin{tikzpicture}
            \tikzstyle{obj}  = [regular polygon,regular polygon sides=8, minimum width=10pt, draw,fill=white, inner sep=0.1cm, font=\scriptsize]
            \tikzstyle{obj1}  = [regular polygon,regular polygon sides=8, minimum width=10pt, draw,fill=green!20, inner sep=0.1cm, font=\scriptsize]
            \tikzstyle{obj3}  = [circle, minimum width=10pt, draw,fill=blue!20, inner sep=0.1cm, font=\scriptsize]
            \tikzstyle{obj2}=[ midway, fill=white, font=\scriptsize]
            \draw (0,0) node[obj1] (1)  {1}
                 ++(0:2.0cm) node[obj1] (2)  {2}
                 ++(288:2.0cm) node[obj1] (3) {3}
                 ++(216:2.0cm) node[obj1] (4)  {4}
                ++(144:2.0cm) node[obj] (5)  {5};
            \path[<->] (1) edge node[obj2] {1} (2);
            \path[<->] (3) edge node[obj2] {1} (4);
        \end{tikzpicture}
\label{fig:figure2}
\end{minipage}
\caption{The weighted encodable graph for a full mesh of size 4, 5 and 6 respectively}
   \label{fig:weighted_encodable_graphFullMesh}
\end{figure*}
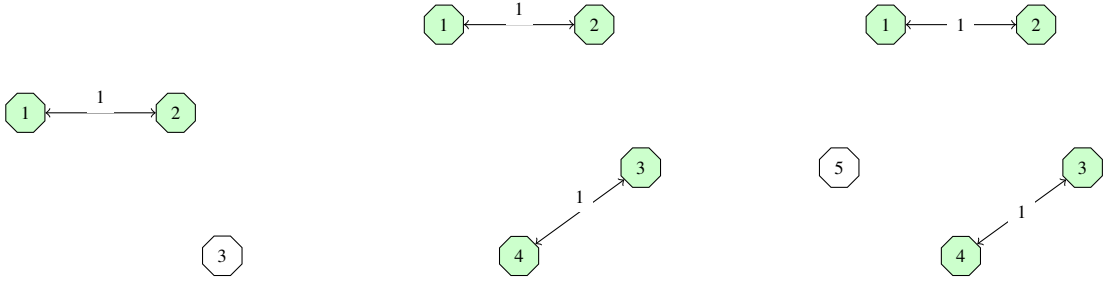

For the network coded approach the network power consumption is given by
\begin{equation}
  P=P_0-(\frac{p_p+p_t}{B})\sum_{d_1,d_2}\sum_{m,n}\min(V_{d_1},V_{d_2})\frac{\beta_{mn}^{d_1 d_2}}{2},
  \label{eq:fullmesh4}  
\end{equation}
which can be reduced to the following, given that equal traffic demands that produces the highest savings
\begin{equation}
  P=P_o- (\frac{p_p+p_t}{B})\frac{V}{2} \sum_{d1,d2}\sum_{m,n}\beta_{mn}^{d_1 d_2}.
  \label{eq:fullmesh4_1}  
\end{equation}

\begin{equation}
  P=P_o- (\frac{p_p+p_t}{B})\frac{V}{2} \sum_{d1,d2}\sum_{m,n}\beta_{mn}^{d_1 d_2}.
  \label{eq:fullmesh5}  
\end{equation}
Since the number of encodable pairs in each cluster in the encodable graph depends on the total number of nodes in the network, the total number of encoded nodes depends on the network size. This is illustrated in Fig. \ref{fig:weighted_encodable_graphFullMesh} for full mesh topologies of size 4 (clusters of size 3), 5 and 6 nodes respectively. If the network has an even number of nodes, then each cluster in the encodable graph will have an odd number of demands (i.e. each receiving node in the network will have demands from $N-1$ nodes, $N$ is even and hence each cluster has an odd number of demands). With an odd number of demands, one demand cannot be paired and hence cannot be network coded and is therefore transmitted using conventional router ports and transponders. This leads to a higher power consumption compared  to a network with an odd number of nodes. In the latter case (network with an odd number of nodes) each cluster has an even number of demands, therefore all demands can be encoded leading to higher power savings. As such the odd and even cases have to be treated separately. 
%Nodes with an even number of nodes have an even number of encodable graph clusters, and with full encoding potential no node will be left out of the cluster. For this reason we divide the derivations into two cases, the odd and even number of nodes.\par
For the full mesh topology with an odd number of nodes (e.g. 5 nodes network, 4 cluster nodes), any two encodable demands have a single hop shared between them (recall the working path for the full mesh is a single hop, and the protection path is 2 hops), therefore
\begin{equation}
\sum_{d1,d2}\sum_{m,n}\beta_{mn}^{d_1 d_2}=N(N-1).
  \label{eq:fullmesh6}  
\end{equation}
Therefore the total power consumption for the network coded case is
\begin{equation}
  P=(\frac{p_p+p_t}{B})N(N-1)\frac{5V}{2}.
  \label{eq:fullmesh6_2}  
\end{equation}
Therefore the total savings is given by
\begin{equation}
  \phi_{odd}=1-\frac{(\frac{p_p+p_t}{B})N(N-1)\frac{5V}{2}}{3(\frac{p_p+p_t}{B})V N(N-1)}=0.166
  \label{eq:fullmesh6_3}  
\end{equation}
which means the savings are upper bounded by a value of 16.67\%. \par 
For the full mesh topology that has an even number of nodes, each cluster will have an odd number of encodable demands, which means that a single encodable node (demand) will not be encoded due to the pairing of all other demands, making the number of encodable demands $N-2$, in each of the $N$ clusters. This fact makes the power savings for the even case less than the power savings of the odd case in (\ref{eq:fullmesh6_3}). With $N$ clusters, and $N-2$ encodable demands in each cluster, the total number of shared hops is given by
\begin{equation}
\sum_{d_1,d_2}\sum_{m,n}\beta_{mn}^{d_1 d_2}=N(N-2).
  \label{eq:fullmesh6_4}  
\end{equation}
The total power consumption of the even case of the full mesh topology under network coding becomes
\begin{equation}
  P=3(\frac{p_p+p_t}{B})VN(N-1)-(\frac{p_p+p_t}{B})V\frac{N(N-2)}{2},
  \label{eq:fullmesh6_5}  
\end{equation}
which gives
\begin{equation}
  P=(\frac{p_p+p_t}{B})VN \left( \frac{5N-4}{2} \right).
  \label{eq:fullmesh6_6}  
\end{equation}
Therefore, the power saving is given by
\begin{equation}
  \phi_{even}=1-\frac{(\frac{p_p+p_t}{B})VN \left( \frac{5N-4}{2}\right)}  {3(\frac{p_p+p_t}{B})V N(N-1)},
  \label{eq:fullmesh7}  
\end{equation}
which leads to
\begin{equation}
  \phi_{even}=\frac{N-2}{6(N-1)}.
  \label{eq:fullmesh8}  
\end{equation}
From equations (\ref{eq:fullmesh6_3}) and (\ref{eq:fullmesh8}), we can see that the power consumption fluctuates between the upper value (i.e. 16.67\%) when the number of nodes is odd,  and the value given by equation (\ref{eq:fullmesh8}) with an even number of nodes. These fluctuations, however, decrease as the number of nodes grows making the network power consumption converge to 16.67\% for any number of nodes. This decrease in fluctuations follows the inverse of the number of nodes and is given by
\begin{equation}
  \epsilon(N)=\frac{1}{6}-\frac{N-2}{6(N-1)}=\frac{1}{6(N-1)},
  \label{eq:fullmesh9}  
\end{equation}
and for a very large number of nodes
\begin{equation}
  \lim_{N\to\infty} \epsilon(N)=\lim_{N\to\infty}\frac{1}{6(N-1)}=0.
  \label{eq:fullmesh10}  
\end{equation}

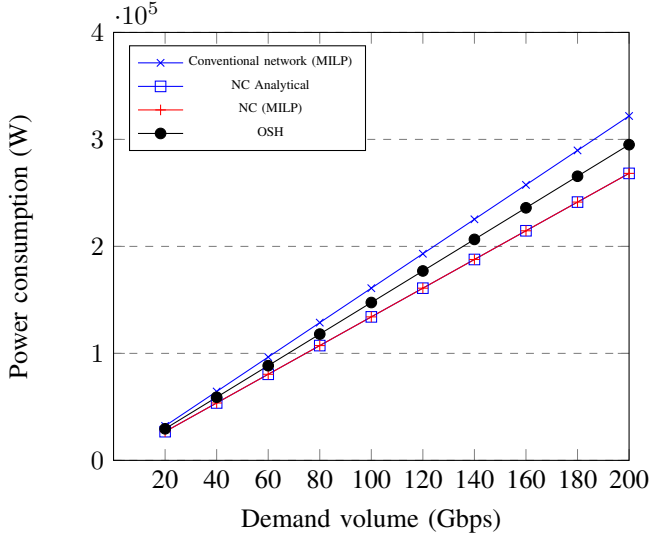
\begin{figure}[h]
    \centering
    \begin{tikzpicture}[scale=1]
\begin{axis}[
    xlabel={Demand volume (Gbps)},
    ylabel={Power consumption (W)},
    xtick=data,
    xmin=0, xmax=5,
    xticklabels={20,40,60,80,100,120,140,160,180,200},
    ymin=0, ymax=400000,
     legend style={legend pos=north west,font=\tiny},
    ymajorgrids=true,
    grid style=dashed,
    ticklabel style = {},
    label style={},
]
 \addplot[
    color=blue,
    mark=x,
    ]
coordinates {(0.5,32190)(1,64380)(1.5,96570)(2,128760)(2.5,160950)(3,193140)(3.5,225330)(4,257520)(4.5,289710)(5,321900)

};
    \addlegendentry{Conventional network (MILP)};
\addplot[
    color=blue,
    mark=square,
    ]
coordinates {(0.5,26825)(1,53650)(1.5,80475)(2,107300)(2.5,134125)(3,160950)(3.5,187775)(4,214600)(4.5,241425)(5,268250)

};
    \addlegendentry{NC Analytical};
    
    \addplot[
    color=red,
    mark=+,
    ]
coordinates {(0.5,26825)(1,53650)(1.5,80475)(2,107300)(2.5,134125)(3,160950)(3.5,187775)(4,214600)(4.5,241425)(5,268250)
};
    \addlegendentry{NC (MILP)};
    
\addplot[
    color=black,
    mark=*,
    ]
coordinates {
 (0.5,  29508)
 (1,    59015)
 (1.5,  88523)
 (2,   118030)
 (2.5, 147540)
 (3,   177045)
 (3.5, 206550)
 (4,   236060)
 (4.5, 265570)
 (5,   295075)
};
    \addlegendentry{OSH};
  
\end{axis}
    \end{tikzpicture}
    \caption{Power consumption of the 5 node full mesh topology with equal demands  using the MILP, Heuristic and analytical bound}
    \label{fig:5nodesEqualMesh}
\end{figure}
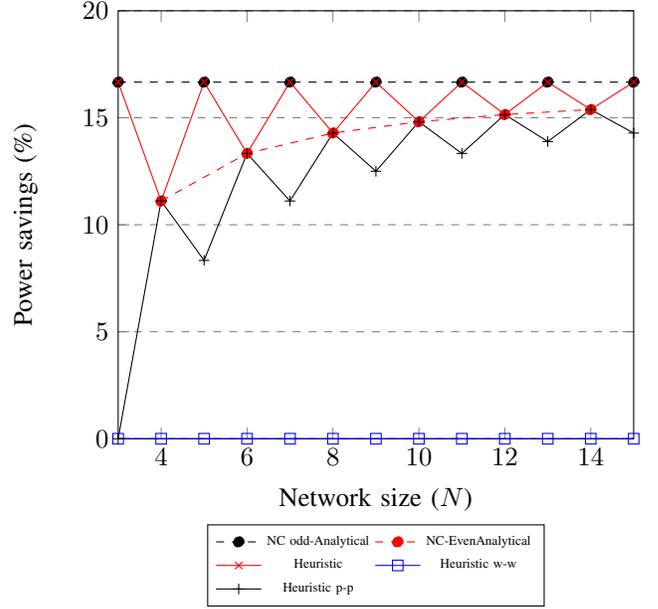
\begin{figure}[h]
    \centering
    \begin{tikzpicture}[scale=1]
\begin{axis}[
    xlabel={Network size ($N$)},
    ylabel={Power savings (\%)},
    xmin=3, xmax=15,
    ymin=0, ymax=20,
     legend style={
			at={(0.5,-0.2)},
			anchor=north,
			legend columns=2, font =\tiny},
     %legend style={at={(0,0)},
%anchor=north east,at={(axis description cs:0,-0.1)}, font=\tiny},
    ymajorgrids=true,
    grid style=dashed,
]

\addplot [domain=3:15,dashed,mark=*, samples=7,unbounded coords=jump]{16.667};

    \addlegendentry{NC odd-Analytical}

 \addplot [color=red,dashed ,mark=*,domain=4:14, samples=6,unbounded coords=jump]{100*(x-2)/(6*(x-1))};

    \addlegendentry{NC-EvenAnalytical}
\addplot[
    color=red,
    mark=x,
    ]
    coordinates {
  (3,  16.6667)
  (4,  11.1111)
  (5,  16.6667)
  (6,  13.3333)
  (7,  16.6667)
  (8,  14.2857)
  (9,  16.6667)
  (10, 14.8148)
  (11, 16.6667)
  (12, 15.1515)
  (13, 16.6667)
  (14, 15.3846)
  (15, 16.6667)
    };
    \addlegendentry{Heuristic}
    
\addplot[
    color=blue,
    mark=square,
    ]
    coordinates {
  (3,   0 )
  (4,   0 )
  (5,   0 )
  (6,   0 )
  (7,   0 )
  (8,   0 )
  (9,   0 )
  (10,  0 )
  (11,  0 )
  (12,  0 )
  (13,  0 )
  (14,  0 )
  (15,  0 )
    };
    \addlegendentry{Heuristic w-w}

%    \addplot[
%    color=black,
%    mark=triangle,
%    dotted,
%    ]
%    coordinates {
%   (3,  16.6667)
%   (4, 11.1111 )
%   (5,  8.3333 )
%   (6,  6.6667 )
%   (7,  5.5556 )
%   (8,  4.7619 )
%   (9,  4.1667 )
%   (10, 3.7037 )
%   (11, 3.3333 )
%   (12, 3.0303 )
%   (13, 2.7778 )
%   (14, 2.5641 )
%   (15, 2.3810 )
%   
%};
%    \addlegendentry{Heuristic w-p}
%    
%    
%        \addplot[
%    color=green,
%    mark=*,
%    opacity=0.5,
%    dotted,
%    ]
%    coordinates {
%   (3,  16.6667)
%   (4, 11.1111 )
%   (5,  8.3333 )
%   (6,  6.6667 )
%   (7,  5.5556 )
%   (8,  4.7619 )
%   (9,  4.1667 )
%   (10, 3.7037 )
%   (11, 3.3333 )
%   (12, 3.0303 )
%   (13, 2.7778 )
%   (14, 2.5641 )
%   (15, 2.3810 )
%	};
%    \addlegendentry{Heuristic p-w}

\addplot[
    color=black,
    mark=+,
    ]
    coordinates {
 (3,  0       )
   (4, 11.1111)
   (5,   8.3333 )
   (6,  13.3333 )
   (7,  11.1111 )
   (8,  14.2857 )
   (9,  12.5000 )
   (10, 14.8148 )
   (11, 13.3333 )
   (12, 15.1515 )
   (13, 13.8889 )
   (14, 15.3846 )
   (15, 14.2857 )
};
    \addlegendentry{Heuristic p-p}

\end{axis}
\end{tikzpicture}
    \caption{The power savings of the full mesh topology under different network sizes}
    \label{fig:MeshSize}
\end{figure}

Fig. \ref{fig:5nodesEqualMesh} Shows a comparison between the power consumption between the MILP, analytical and the OSH heuristic for the 5 nodes full mesh topology and compares them with the conventional MILP scenario. It clearly shows that the analytical results matches the MILP results. It also shows a linear dependency between the power consumption and the average demand volume, as can be seen from equation  (\ref{eq:fullmesh6_2}) when the number of nodes $N$ is fixed for a given network. This slope of the curve is given as $(\frac{p_p+p_t}{B}\frac{5}{2}N(N-1))$
\\ 
We show in Fig. \ref{fig:MeshSize} the power savings of full mesh topologies with a number of nodes ranging from 3 nodes up to 15 nodes. The concept of multi path protection and therefore the concept of network coded protection doesn’t apply to networks with number of nodes below 3.  It is obvious that encoding both working flows together produces no savings as both working flows use the direct link between each node in the network which is not shared with the direct link of a working path of another demand. It also shows that the Optimal Search Heuristic (OSH) \cite{musa2017energy} is superior,  while the form of the heuristic that encodes  protection paths together produces optimal savings at even network sizes. 
%The figure also shows that at odd number of nodes, the total power savings is the sum of the power savings of the heuristic with both protection paths encoded and the result of the heuristic with a protection and a working path encoding.
The savings of the  optimal heuristic jumps increasing and decreasing as the network size changes between odd and even number of nodes, agreeing with the analytical formulas, but overall converges to the maximum possible savings value (i.e. 16.67\%).\par

%The figure also shows that both second and third approaches of encoding working and protection flows together have the exact same savings and converges to zero savings as the network size increases, and have the highest of savings at the smallest possible considerable network of size 3 (i.e. 16.67\%). \par

\subsection{Ring topology}
The power consumption of the conventional 1+1 protection of the ring is given as
\begin{equation}
P_0=(\frac{p_p+p_t}{B})\left(\sum_{d\in D}\sum_{m,n}V^{d}(x_{mn}^{d}+y_{mn}^{d})\right).
\label{eq:ring1}
\end{equation}

The total count of working hops for the odd number of nodes is given as
\begin{multline}
h_w=\sum_{d \in D}\sum_{m,n}x_{mn}^d=2N\left(1+2+...+(\frac{N-1}{2}) \right)\\=\frac{(N-1)N(N+1)}{4}.
\label{eq:ring1}
\end{multline}

Since each working path of length $k$ has a protection path of length $N-k$ in the other direction, this makes the total number of protection hops for the case of odd number of nodes 

\begin{multline}
h_p=\sum_{d \in D}\sum_{m,n}y_{mn}^d\\=2N\left( N-1+N-2+...+N-\frac{N-1}{2} \right) \\=N(N-1)(\frac{3N-1}{4}),
\end{multline}

and the total number of hops of both working and protection paths for the odd number of nodes is given as
\begin{multline}
\sum_{d\in D}\sum_{m,n}(x_{mn}^{d}+y_{mn}^{d})\\=\frac{(N-1)N(N+1)}{4}+N(N-1)(\frac{3N-1}{4})\\=N^3-N^2.
\label{eq:oddConv}
\end{multline}

For the case of even number of nodes, the number of working hops is
\begin{multline}
h_w=\sum_{d \in D}\sum_{m,n}x_{mn}^d\\
=2N\left(1+2+...+(\frac{N-2}{2}) \right)+N \frac{N}{2}\\
=\frac{N^3}{4}, 
\label{eq:ring3}
\end{multline}
and the number of protection hops is given by
\begin{multline}
h_p=\sum_{d \in D}\sum_{m,n}y_{mn}^d\\
=2N\left( N-1+N-2+...+N-\frac{N-2}{2} \right)+N \frac{N}{2} 
\end{multline}
\begin{multline}
h_p=2N\left(N(\frac{N-2}{2})-1-2-...-\frac{N-2}{2}\right)+\frac{N^2}{2}\\
=\frac{N^2(3N-4)}{4}
\end{multline}
and the total number of hops of both working and protection paths for the even number of nodes is given as
\begin{equation}
\sum_{d\in D}\sum_{m,n}(x_{mn}^{d}+y_{mn}^{d})=\frac{N^3}{4}+\frac{N^2(3N-4)}{4}=N^3-N^2
\label{eq:evenConv}
\end{equation}
This expression is for the conventional case and is the same for rings of odd and even number of nodes (i.e. (\ref{eq:oddConv}) is the same as (\ref{eq:evenConv})). 
\subsubsection{Rings with odd size}
\begin{figure}[h]
    \includegraphics[scale=0.475]{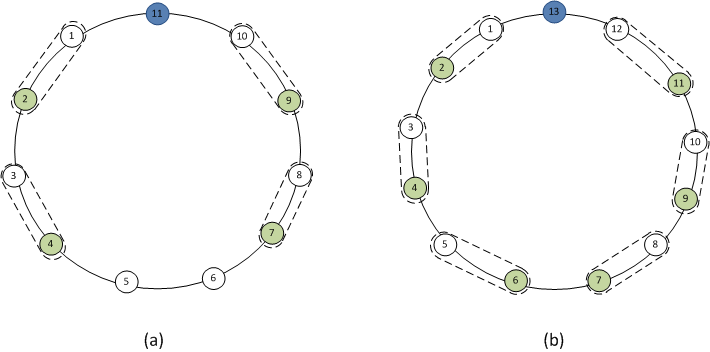}
    \caption{Encoding pairs for two rings with different odd number of nodes}
    \label{fig:RingOdd2}
\end{figure}

We start with the case of a ring with odd number of nodes, as shown in Fig. (\ref{fig:RingOdd2}). The figure shows a ring with $11$ and $13$ nodes where all nodes send to node $11$ and $13$, respectively. To maximise the number of shared links, protection paths are encoded together so the longest protection path is encoded with the second longest protection path and so on, leading to a number of shared hops that is equal to the number of hops of the shorter protection path. This is shown in Fig.  (\ref{fig:RingOdd2}), where we pair the source nodes of demands that can be encoded. Fig.  \ref{fig:RingOdd2} shows that the demands $(1,11)$ and $(2,11)$ are encoded together, where demand $(1,11)$ has a protection path with a length of $10$ hops and demand $(2,11)$ has a protection path of $9$ hops, leading to $9$ shared hops. The same principle applies between demands $(3,11)$ and $(4,11)$ leading to $7$ shared hops, which is equal to the length of the protection path of demand $(4,11)$. The same applies to  demands $[(10,11), (9,11)]$ and $[(8,11), (7,11)]$. As node $5$, and node $6$ do not share a protection path because they send their protection signals in opposite directions, they do not get encoded together.\par 
The second example of a 13 nodes ring, shows that all nodes can find another node to be paired with. Therefore, compared to the $11$ nodes ring, better savings are achieved. As a result, the power savings obtained under network coding go up and down as the number of nodes in the odd ring changes between the odd number where $\frac{N-1}{2} \mod 2 =1$, classified as odd-1, and the odd number where $\frac{N-1}{2} \mod 2 =0$, which is classified as odd-2. For example, when $N=11$, we have $\frac{11-1}{2} \mod 2=1$ meaning $11$ nodes belong to group 1 (i.e. odd-1), and when $N=13$, we have $\frac{13-1}{2} \mod 2=0$ meaning a ring with $13$ nodes belongs to group 2 (i.e. odd-2). \par

We start with the first odd group (i.e. odd-1), of which Fig. \ref{fig:RingOdd2} (a) is an example, with $11$ nodes. To calculate the total number of shared hops between all encoded demands (i.e. the hop count of green nodes as explained earlier), we first determine the total number of shared hops between encoded demands destined to one destination (i.e. to destination node 11 in the example in Fig. \ref{fig:RingOdd2}a), and then multiply it by the total number of destinations (i.e. $N$). For demands going to the same destination, it can be seen that for each demand on one side of the destination node there exists another demand with the same length of the protection path on the other side of the ring. Therefore we derive an expression for one side of the ring and then multiply the result by two.
To determine the number of shared hops on one side of the ring, we calculate the number of pairs on that side, which is given as $\frac{N-3}{4}$, deduced by removing 
three nodes (i.e.  the destination node (node $11$), and the other two non-encodable nodes ($5$) and ($6$)), then dividing by two to account for one side, and dividing again by two to count the pairs on that side. 
%the length of the protection path for encoded demands follows the following sequence, N-2, N-4, ...etc for N-3/4 terms.

%we divide the total demands by their destination nodes, in the current case node $11$ is the destination node, and therefore we calculate the total number of hops for a single destination and multiply by $N$ to generalize for the whole set of demands. The ring is divided further into two symmetrical halves (one containing $(1,2,3,4)$ and the other contains ($7,8,9,10$) as encodable demand sources). Again, due to symmetry, the total hop count will be twice the total hop count of one half, and hence multiplied by $2N$ to cover the two halves and the whole demands set.\par
%The number of terms that get encoded (number of green nodes in each half in Figure \ref{fig:RingOdd2}a) is $\frac{N-3}{4}$ which is deduced by removing three nodes (i.e.  the destination node (node $11$), and the other two non-encodable nodes ($5$) and ($6$)), then dividing by $2$ to produce one half, and dividing by $2$ again to count the pairs in a given half. This is a general case for all ring with odd sizes belonging to the first set. For each green node, the length of the protection path is the the total number of nodes minus the length of the working path. 
%The expression inside the square brackets represents the sum ($9+7$) in this example, converted into $(11-2)+(11-4)$,  that equals $2*11-2*(1+2)$. This is generalized for any number of nodes (here $N=11$) and any number of encodable terms $\frac{N-1}{4}$ which equals $2$ in our case. \par 

Therefore, the total number of shared hops for the odd-1 ring  group is

\begin{equation}
h_t(odd1)=2N \bigg[(N-2)+(N-4)+...+(N-\frac{N-3}{2}) \bigg], 
\label{eq:ringodd1_1}
\end{equation}
Equation \ref{eq:ringodd1_1} can be described with the aid of Fig. \ref{fig:RingOdd2}a, where on one side, the encoded pair (node $1$ and node $2$) have a shared hop count of $(N-2)$, which is added to $(N-4)$ that represents the shared hop count between the encoded pair (node $3$ and node  $4$). For larger rings, the number of shared hops continues to decrease by two, and the final term is given by $(N-\frac{N-3}{2})$. After adding similar terms, we have

\begin{equation}
h_t(odd1)=2N \bigg[N (\frac{N-3}{4})-2-4-...-\frac{N-3}{2} \bigg], 
\label{eq:ringodd1_2}
\end{equation}
and that gives
\begin{equation}
h_t(odd1)=2N \bigg[N(\frac{N-3}{4})-2 \sum_{k=1}^{\frac{N-3}{4}}k \bigg]=\frac{N(N-3)(3N-1)}{8}.
\label{eq:ringodd1}
\end{equation}

%The expression components in equation (\ref{eq:ringodd1}) corresponds to the first set of rings with odd number of nodes, of which the ring of Figure \ref{fig:RingOdd2}(a) is an example.  The ring is divided by destination nodes (current ring has node $11$ as a destination), and further divided into two symmetrical halves (one containing nodes $(1,2,3,4)$ and the other contains $7,8,9,10$ of encodable nodes). The derivation will focus on one half of a ring, and on one destination nodes, and is generalized by multiplying by $2N$, where the 2 covers both ring halves and the $N$ covers all destination nodes. 

The total power saving for this case is represented by
\begin{equation}
\phi_{odd1}=\frac{N(N-3)(3N-1)}{8(N^3-N^2)}.
\end{equation}

\begin{equation}
\lim_{N \to \infty }\phi_{odd1}=\lim_{N \to \infty }\frac{N(N-3)(3N-1)}{8(N^3-N^2)}=\frac{3}{8}=37.5\%.
\end{equation}

For the second odd group, represented by Fig.  (\ref{fig:RingOdd2}b), the total number of encodable demands in each half is given by $\frac{N-1}{4}$ as only the destination node is not selected. This gives
\begin{equation}
h_t(odd2)=2N \bigg[(N-2)+(N-4)+...+(N-\frac{N-1}{2}) \bigg], 
\end{equation}
which gives 
\begin{equation}
h_t(odd2)=2N \bigg[N \frac{N-1}{4} -2-4-\frac{N-1}{2} \bigg], 
\end{equation}
which can be written as
\begin{equation}
h_t(odd2)=2N \bigg[N\frac{N-1}{4}-2 \sum_{k=1}^{\frac{N-1}{4}}k \bigg]=\frac{3}{8}N(N-1)^2.
\end{equation}
This makes the total power saving
\begin{equation}
\phi_{odd2}=\frac{\frac{3}{8}N(N-1)^2}{N^3-N^2},
\end{equation}

\begin{equation}
\lim_{n \to \infty }\phi_{odd2}=\lim_{n \to \infty }\frac{\frac{3}{8}N(N-1)^2}{N^3-N^2}=\frac{3}{8}=37.5\%.
\end{equation}

\subsubsection{Rings with even sizes}
\begin{figure}[h]
    \includegraphics[scale=0.475]{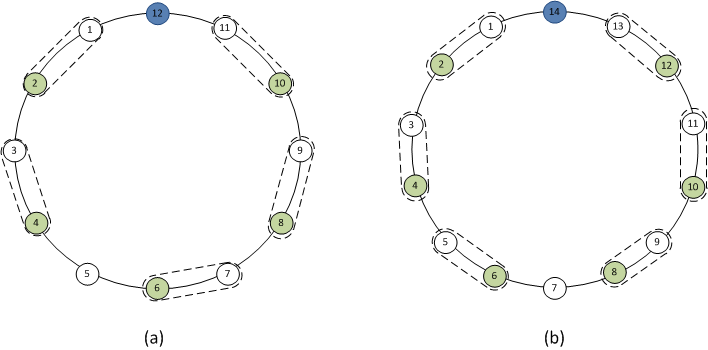}
    \caption{Encoding pairs for two rings with different even number of nodes}
    \label{fig:RingEven2}
\end{figure}

Here we also face the same distinction between two sets of even ring sizes, where rings of size ($4,8,12,...$) will be in a different group (i.e. even-1) and have a different expression compared to the group (i.e. even-2) containing the other ring sizes ($6,10,14,...$). This is illustrated in Fig. (\ref{fig:RingEven2}). In both cases, the destination node and another demand source node (i.e. node 5 in Fig.  \ref{fig:RingEven2}a and node 7 in Fig.  \ref{fig:RingEven2}b) are not paired. A ring with an even size $N$ is classified into its appropriate group, and hence its bound, by checking  if $\frac{N-2}{2} \mod 2=1$, if so it belongs to the even-1 group, and it belongs to even-2 when $\frac{N-2}{2} \mod 2=0$. For example, when $N=12$, $\frac{12-2}{2} \mod 2=1$ meaning $12$ nodes belong to group 1, and when $N=14$, $\frac{14-2}{2} \mod 2=0$ meaning a ring with $14$ nodes belong to group 2. \par
%When the ring size is $12$, two nodes get left out of encoding, two encodings at each half of the ring manifest themselves, in addition to the central node encoding. On the other hand, a ring with size $14$ manages to encode 3 pairs on each side having left also two nodes off encoding.\par
%We derive expressions for the two classes of rings with odd number of nodes, and given a number of nodes $N$ the certain formula and group is determined by $\frac{N-2}{2} \mod 2$. For example, when $N=12$, $\frac{12-2}{2} \mod 2=1$ meaning $12$ nodes belong to group 1, and when $N=14$, $\frac{14-2}{2} \mod 2=0$ meaning a ring with $14$ nodes belong to group 2. \par

We start by the even-1 group represented by Fig.  \ref{fig:RingEven2}a.
%The total number of shared hops will follow the shape $N(2z+\frac{N}{2})$, where $N$ is the generalisation for N destinations, $z$ is the number of shared links on one side of the ring, doubled to cover the whole ring, and $\frac{N}{2}$ is the shared hop count for the additional encoded pair (i.e. demand pair with source nodes $6$ and $7$). 
The total number of encoded pairs in the ring is given by $N(2\frac{N-4}{4}+\frac{N}{2})$, where the number of encoded pairs on one side is $\frac{N-4}{4}$ which is deduced by removing four nodes (i.e. 12, 5, 6 and 7) to maintain the symmetry needed for the whole expression, while  the  $\frac{N}{2}$ accounts for the shared hop count of the encoded pair (node $6$ and node $7$).
%The value $z$ in this example (12 nodes) equals $10+8$ for the sum of shared protection paths of the two pairs in each half. This can be rewritten to simplify derivations to  $10+8=(6+4)+(6+2)=6+2+6+4=\frac{N}{2}+\frac{N}{2}+2(1+2)$, where the factor $2$ represents the number of encoded pairs in each half, given by removing 4 nodes (12, 5, 6, and 7), represented by $\frac{N-4}{4}$. 
Therefore the total number of shared hops is given by

%\begin{equation}
%h_t(even1)=N \left( 2\big[(\frac{N}{2}+2)+(\frac{N}{2}+4)+...+(\frac{N}{2}+(\frac{N-4}{2}))\big]+\frac{N}{2} \right)
%\end{equation}
%The expression can be represented as:
\begin{equation}
h_t(even1)=N \left( 2\bigg[N\frac{N-4}{4} -2-4-...-\frac{N-4}{2}\bigg]+\frac{N}{2} \right), 
\end{equation}
The additional $\frac{N}{2}$ term inside the brackets is the shared hop count as a result of  encoding between node $6$ and node $7$) in Fig. \ref{fig:RingEven2}a.  
%\begin{equation}
%h_t(even1)=N \left( 2\bigg[N\frac{N-4}{2} -2-4-...-\frac{N-4}{4}\bigg]+\frac{N}{2} \right), 
%\end{equation}

%and that gives
%\begin{equation}
%h_t(even1)=N \left( 2\bigg[N\frac{N-4}{4} -2\left( %1+2+...+\frac{N-4}{4}\right) \bigg]+\frac{N}{2} \right),
%\end{equation}
which equals
\begin{equation}
h_t(even1)=N \left( 2\big[N\frac{N-4}{4} -2 \sum_{k=1}^{\frac{N-4}{4}}k\big]+\frac{N}{2} \right)
\end{equation}
which gives

\begin{equation}
h_t(even1)=\frac{N^2}{2} \big[1+\frac{3(N-4)}{4} \big]. 
\end{equation}

The savings for the even-1 ring is 

\begin{equation}
\phi_{even1}=\frac{\frac{1}{2}N^2(\frac{3N}{4}-2)}{N^3-N^2},
\end{equation}
therefore
\begin{equation}
\lim_{N \to \infty }\phi_{even_1}=\lim_{N \to \infty }\frac{\frac{1}{2}N^2(\frac{3N}{4}-2)}{N^3-N^2}=\frac{3}{8}=37.5\%.
\end{equation}

For the even-2 ring, the same approach applies, just by removing two nodes (destination node and central node), and it becomes completely symmetrical, having a number of encodable demands on each side of the destination node given by $\frac{N-2}{4}$. Therefore giving the following total number of shared hops

%\begin{equation}
%h_t(even2)=N \left( 2\big[(\frac{N}{2}+1)+(\frac{N}{2}+3)+...+(\frac{N}{2}+(\frac{N-4}{2}))\big] \right),
%\end{equation}
%which gives
\begin{equation}
h_t(even2)=N \left( 2\bigg[N\frac{N-2}{4} -2-4-...-\frac{N-2}{2}\bigg] \right), 
\end{equation}
which gives

\begin{equation}
h_t(even2)=N \left( 2\bigg[N\frac{N-2}{4}- 2\sum_{k=1}^{\frac{N-2}{4}}k\bigg] \right), 
\end{equation}
which can be written as
\begin{equation}
h_t(even2)=\frac{N(N-2)(3N-2)}{8}.
\end{equation}

so the savings of the even-2 ring is
\begin{equation}
\phi_{even2}=\frac{\frac{1}{8}N(N-2)(3N-2)}{N^3-N^2}, 
\end{equation}
therefore
\begin{equation}
\lim_{N \to \infty }\phi_{even_2}=\lim_{N \to \infty }\frac{\frac{1}{8}N(N-2)(3N-2)}{N^3-N^2}=\frac{3}{8}=37.5\%.
\end{equation}

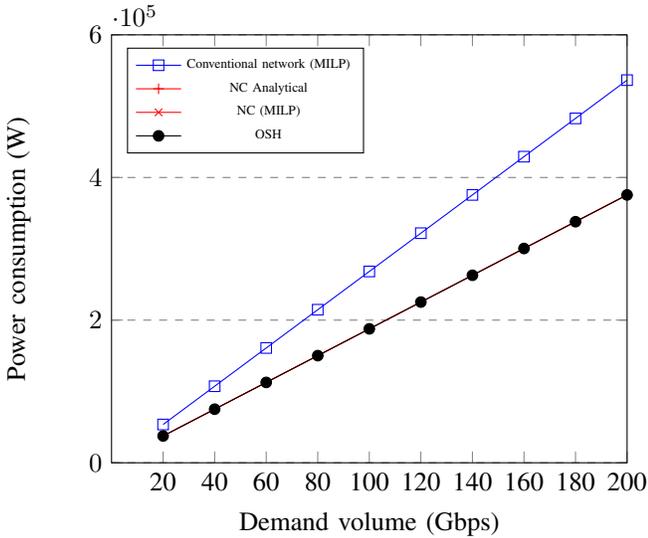
\begin{figure}[h]
    \centering
    \begin{tikzpicture}[scale=1]
\begin{axis}[
    xlabel={Demand volume (Gbps)},
    ylabel={Power consumption (W)},
    xtick=data,
    xmin=0, xmax=5,
    xticklabels={20,40,60,80,100,120,140,160,180,200},
    ymin=0, ymax=600000,
     legend style={legend pos=north west,font=\tiny},
    ymajorgrids=true,
    grid style=dashed,
    ticklabel style = {},
    label style={},
]
 
\addplot[
    color=blue,
    mark=square,
    ]
coordinates {(0.5,53650)(1,107300)(1.5,160950)(2,214600)(2.5,268250)(3,321900)(3.5,375550)(4,429200)(4.5,482850)(5,536500)

};
    \addlegendentry{Conventional network (MILP)};
    \addplot[
    color=red,
    mark=+,
    ]
coordinates {(0.5,37555)(1,75110)(1.5,112665)(2,150220)(2.5,187775)(3,225330)(3.5,262885)(4,300440)(4.5,337995)(5,375550)
};
    \addlegendentry{NC Analytical};
    \addplot[
    color=red,
    mark=x,
    ]
coordinates {(0.5,37555)(1,75110)(1.5,112665)(2,150220)(2.5,187775)(3,225330)(3.5,262885)(4,300440)(4.5,337995)(5,375550)
};
    \addlegendentry{NC (MILP)};
    
\addplot[
    color=black,
    mark=*,
    ]
coordinates {(0.5,37555)(1,75110)(1.5,112665)(2,150220)(2.5,187775)(3,225330)(3.5,262885)(4,300440)(4.5,337995)(5,375550)

};
    \addlegendentry{OSH};
  
\end{axis}
    \end{tikzpicture}
    \caption{Power consumption of the 5 node ring topology with equal demands  using the MILP, Heuristic and analytical bound}
    \label{fig:5nodesEqualRing}
\end{figure}

Fig. \ref{fig:5nodesEqualRing} Shows a comparison between the power consumption between the MILP, analytical and the OSH heuristic for the 5 nodes ring topology and compares them with the conventional MILP scenario, demonstrating an exact match between the analytic and MILP results. 

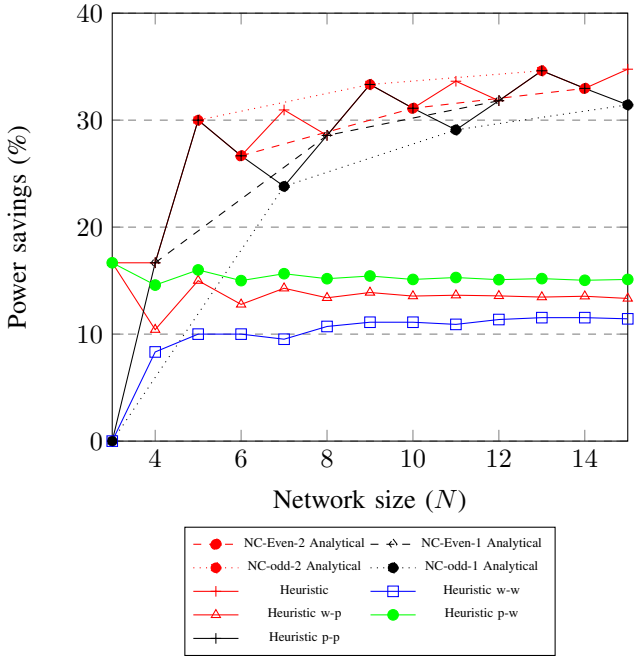
\begin{figure}[h]
    \centering
    \begin{tikzpicture}[scale=1]
\begin{axis}[
    xlabel={Network size ($N$)},
    ylabel={Power savings (\%)},
    xmin=3, xmax=15,
    ymin=0, ymax=40,
          legend style={
			at={(0.5,-0.2)},
			anchor=north,
			legend columns=2, font =\tiny},
    ymajorgrids=true,
    grid style=dashed,
]
  \addplot [color=red, mark=*, dashed ,domain=6:14, samples=3,unbounded coords=jump]{(100/8)*((x-2)*(3*x-2))/(x*(x-1))};

    \addlegendentry{NC-Even-2 Analytical}
    
\addplot [color=black,dashed ,mark=diamond ,domain=4:12, samples=3,unbounded coords=jump]{(100/8)*(3*x-8)/(x-1)};

    \addlegendentry{NC-Even-1 Analytical}

\addplot [color=red,dotted,mark=* ,domain=5:13, samples=3,unbounded coords=jump]{(100*(3/8)*(x-1)/x};

    \addlegendentry{NC-odd-2 Analytical}
    
 \addplot [color=black,dotted,mark=* ,domain=3:15, samples=4,unbounded coords=jump]{(100*(1/8)*(x-3)*(3*x-1)/(x*(x-1))};

    \addlegendentry{NC-odd-1 Analytical}  
 \addplot[
    color=red,
    mark=+,
    ]
    coordinates {
  (3,   16.6667 )
  (4,   16.6667 )
  (5,   30.0000 )
  (6,   26.6667 )
  (7,   30.9524 )
  (8,   28.5714 )
  (9,   33.3333 )
  (10,  31.1111 )
  (11,  33.6364 )
   (12, 31.8182 )
   (13, 34.6154 )
   (14, 32.9670 )
   (15, 34.7619 )
   };
    \addlegendentry{Heuristic}

\addplot[
    color=blue,
    mark=square,
    ]
    coordinates {
  (3,       0)
   (4,  8.3333)
   (5, 10.0000)
   (6, 10.0000)
   (7,  9.5238)
   (8, 10.7143)
   (9, 11.1111)
   (10,11.1111)
   (11,10.9091)
   (12,11.3636)
   (13,11.5385)
   (14,11.5385)
   (15,11.4286)
    };
    \addlegendentry{Heuristic w-w}

    \addplot[
    color=red,
    mark=triangle,
    ]
    coordinates {
   (3, 16.6667)
   (4, 10.4167)
   (5, 15.0000)
   (6, 12.7778)
   (7, 14.2857)
   (8, 13.3929)
   (9, 13.8889)
   (10,13.5556)
   (11,13.6364)
   (12,13.5732)
   (13,13.4615)
   (14,13.5400)
   (15,13.3333)
};
    \addlegendentry{Heuristic w-p}

        \addplot[
    color=green,
    mark=*,
    ]
    coordinates {
   (3,  16.6667)
   (4, 14.5833 )
   (5, 16.0000 )
   (6, 15.0000 )
   (7, 15.6463 )
   (8, 15.1786 )
   (9, 15.4321 )
   (10,15.1111 )
   (11,15.2893 )
   (12,15.0884 )
   (13,15.1874 )
   (14,15.0314 )
   (15,15.1111 )
	};
    \addlegendentry{Heuristic p-w}

\addplot[
    color=black,
    mark=+,
    ]
    coordinates {
   (3,       0)
   (4, 16.6667)
   (5, 30.0000)
   (6, 26.6667)
   (7, 23.8095)
   (8, 28.5714)
   (9, 33.3333)
   (10,31.1111)
   (11,29.0909)
   (12,31.8182)
   (13,34.6154)
   (14,32.9670)
   (15,31.4286)
};
    \addlegendentry{Heuristic p-p}

\end{axis}
\end{tikzpicture}
    \caption{The power savings of the ring topology under different network sizes}
    \label{fig:ringSize}
\end{figure}

We evaluate the impact of the ring size by showing the power savings of ring topologies ranging from 3 nodes up to 15 nodes as shown in Fig. \ref{fig:ringSize}. The figure shows that the analytical formulas developed for the 2 cases of the even number of nodes and the other two cases of the odd number of nodes matches exactly the results of the heuristic, all together converging to the highest possible savings of 37.5\% as the number of nodes grows. The figure also shows that the other heuristics (i.e. w-w, w-p, and  p-w), have comparable savings around 15\% that are far inferior to the OSH heuristic and the p-p heuristic.  \par
The figure also shows that the difference between the values of the analytical formulas for the odd-1 and odd-2 case are higher than the difference of the analytical value between even-1  and even-2 cases. This can be explained with the aid of Fig.  (\ref{fig:RingOdd2}) and Fig.  (\ref{fig:RingEven2}), where in the case of an even number of nodes, 2 nodes get left out each time for both even cases, while for the odd case, one node gets left out in one case and three at the other. This also explains why the first odd group has the highest savings where only 1 node gets left out (all nodes are encoded). It also shows that the power savings of the OSH heuristic are higher than the heuristic p-p in the odd-2 case (e.g. size 7, 11 and 15), because in this case not all nodes are encodable and the heuristic tries all possible combinations while the heuristic p-p chooses only protection paths.  

\section{conclusion}
In this work we developed analytical bounds and closed form expressions for energy efficient survivable IP over WDM networks that use network coding, encoding the protection paths of demands using simple XOR operations. The analytical bounds were developed also for the conventional 1+1 protection scheme without network coding as a function of the average demand volume, the network size and the minimum hop count. We introduced a new concept, the characteristic hop count, and showed that the power consumption of the network coded case is a function of this characteristic hop count alongside the average traffic volume and the network size. We also studied regular topologies with emphasis on the full mesh and the ring topologies, providing a study on large network sizes, proving that the mesh, and ring topologies to exhibit savings that approach asymptotically 16.7\% and 37.5\% respectively. We also provided a closed form expression of the total number of hops as a function of network size for the full mesh and ring topologies.The implementation of network coding in this work uses the same algorithms of routing and path allocation algorithms used in the conventional approach, making the application to existing network a matter of a small incremental addition, as using a simple xor operation is much simpler compared to the highly complex existing techniques such as forward error correction. An interesting direction of further study however is to analyses the additional efficiency gained against the higher complexity incurred by using much higher order network coding techniques. The impact of a cenrtalized control and management using SDN as opposed to the distributed control and the contrast between them regarding the power consumption at different types of codes is also a future direction with significant value. 
\section*{Acknowledgments}
The authors would like to acknowledge funding from the Engineering and Physical Sciences Research Council (EPSRC), INTERNET (EP/H040536/1) and STAR (EP/K016873/1). All data are provided in full in the results section of this paper.

\ifCLASSOPTIONcaptionsoff
  \newpage
\fi



%\bibliographystyle{IEEEtran}
%\printbibliography
%\bibliography{/references.bib}
%\bibliography{NetworkCoding.bib}

\begin{thebibliography}{24}
 
\bibitem{Ahlswede2000}
R. Ahlswede, C. Ning, S. Y. R. Li, and R. W. Yeung, “\emph{Network
information flow,” Information Theory}, IEEE Transactions on, vol. 46,
no. 4, pp. 1204-1216, 2000.

\bibitem{kamal2014network}
A. E. Kamal and M. Mohandespour, \emph{Network coding-based protection,}
Optical Switching and Networking, vol. 11, pp. 189-201, 2014.

\bibitem{babarczi2013realization}
P. Babarczi, G. Biczok, H. Øverby, J. Tapolcai, and P. Soproni, “\emph{Realization strategies of dedicated path protection: A bandwidth cost
perspective,}” Computer Networks, vol. 57, no. 9, pp. 1974-1990, 2013

\bibitem{overby2012cost}
H. Overby, G. Biczok, P. Babarczi, and J. Tapolcai, \emph{Cost comparison ´
of 1+1 path protection schemes: A case for coding,} in Communications
(ICC), 2012 IEEE International Conference on. IEEE, 2012.
\bibitem{musa2015energy}
M. Musa, T. E. El-Gorashi, and J. M. Elmirghani, “\emph{Energy efficient
core networks using network coding,}” in 2015 17th International Conference
on Transparent Optical Networks (ICTON). IEEE, 2015, pp.
1–4.
\bibitem{musa2016energy}
M. Musa, T. E. El-Gorashi, and J. M. Elmirghani, “\emph{Network coding for
energy efficiency in bypass ip/wdm networks,}” in 2016 18th International
Conference on Transparent Optical Networks (ICTON). IEEE,
2016.
\bibitem{kamal2011efficient}
A. E. Kamal and O. Al-Kofahi, “\emph{Efficient and agile 1+ n protection,}”
Communications, IEEE Transactions on, vol. 59, no. 1, pp. 169-180, 2011.
\bibitem{kamal2010network}
A. E. Kamal, “\emph{Network protection for mesh networks: network codingbased
protection using p-cycles,}” Networking, IEEE/ACM Transactions on, vol. 18, no. 1, pp. 67-80, 2010.
\bibitem{aly2008network}
S. A. Aly and A. E. Kamal, “ \emph{ Network protection codes against link failures
using network coding,}” in Global Telecommunications Conference,
2008. IEEE GLOBECOM 2008. IEEE. IEEE, 2008, pp. 1-6.
\bibitem{aly2009network}
Aly, Salah A., and Ahmed E. Kamal, “\emph{Network coding-based protection strategy against node failures,}”
in Communications, 2009. ICC’09. IEEE International Conference on.
IEEE, 2009, pp. 1-5.
\bibitem{barla2010network}
I. B. Barla, F. Rambach, D. A. Schupke, and M. Thakur, “\emph{Network
coding for protection against multiple link failures in multi-domain networks,}”
in Communications (ICC), 2010 IEEE International Conference
on. IEEE, 2010, pp. 1-6

\bibitem{muktadir2012optimum}
A. Muktadir, A. A. Jose, and E. Oki, “\emph{An optimum mathematical
programming model for network-coding based routing with 1+ 1
path protection,}” in World Telecommunications Congress (WTC), 2012.
IEEE, 2012, pp. 1-5.
\bibitem{idzikowski2016survey}
F. Idzikowski, L. Chiaraviglio, A. Cianfrani, J. L. Vizca´ıno, M. Polverini,
and Y. Ye, “\emph{A survey on energy-aware design and operation of core networks,}” IEEE Communications Surveys and Tutorials, vol. 18, no. 2,
pp. 1453-1499, 2016.
\bibitem{dharmaweera2015toward}
M. N. Dharmaweera, R. Parthiban, and Y. A. S¸ ekercioglu, “\emph{Toward a power-efficient backbone network: the state of research,}” IEEE Communications
Surveys and Tutorials, vol. 17, no. 1, pp. 198-227, 2015.
\bibitem{dong2011ip}
X. Dong, T. El-Gorashi, and J. M. Elmirghani, “ \emph{Ip over wdm networks
employing renewable energy sources,}” Lightwave Technology, Journal
of, vol. 29, no. 1, pp. 3-14, 2011.
\bibitem{Dong2011}
X. Dong, T. El-Gorashi, and J. M. H. Elmirghani, “\emph{ Green IP Over
WDM Networks With Data Centers,}” Journal of Lightwave Technology,
vol. 29, no. 12, pp. 1861-1880, Jun. 2011. 

\bibitem{Dong2012}
X. Dong, T. E. H. El-gorashi, and J. M. H. Elmirghani, “\emph{On the Energy Efficiency of Physical Topology Design for IP Over WDM Networks,}”
vol. 30, no. 12, pp. 1931-1942, 2012.
\bibitem{LaweyCloud}
A. Lawey, T. El-Gorashi, and J. Elmirghani, “\emph{ Distributed energy efficient
clouds over core networks,}” Lightwave Technology, Journal of, vol. 32,
no. 7, pp. 1261-1281, April 2014.
\bibitem{Niema2014}
N. Osman, T. El-Gorashi, L. Krug, and J. Elmirghani, “\emph{Energy-efficient
future high-definition TV,}” Lightwave Technology, Journal of, vol. 32,
no. 13, pp. 2364-2381, July 2014.
\bibitem{Lawey2014}
A. Lawey, T. El-Gorashi, and J. Elmirghani, “\emph{Bittorrent content distribution
in optical networks,}” Lightwave Technology, Journal of, vol. 32,
no. 21, pp. 4209-4225, Nov 2014.
\bibitem{Nonde2015}
L. Nonde, T. El-Gorashi, and J. Elmirghani, “\emph{Energy efficient virtual
network embedding for cloud networks,}” Lightwave Technology, Journal
of, vol. 33, no. 9, pp. 1828-1849, May 2015.
\bibitem{MusaJLT}
M. Musa, T. El-Gorashi, and J. Elmirghani, “\emph{Energy efficient routing and network coding assignment in core
networks,}” to be submitted to Journal of Lightwave Technology, IEEE, 2017.
\bibitem{musa2017energy}
M. Musa, T. Elgorashi, and J. Elmirghani, “\emph{Energy efficient survivable
ip-over-wdm networks with network coding,}” Journal of Optical Communications
and Networking, IEEE, vol. 9, no. 3, pp. 207-217, 2017.
\bibitem{Akhilesh2016}
Thyagaturu, Akhilesh S and Mercian, Anu and McGarry, Michael P and Reisslein, Martin and Kellerer, Wolfgang, “\emph{Software defined optical networks (SDONs): A comprehensive survey,}” IEEE Communications Surveys \& Tutorials 18.4 (2016): 2738-2786. 

\bibitem{capone2015detour}
Capone, Antonio and Cascone, Carmelo and Nguyen, Alessandro QT and Sanso, Brunilde, “\emph{Detour planning for fast and reliable failure recovery in SDN with OpenState,}” Design of Reliable Communication Networks (DRCN), 2015 11th International Conference on the. IEEE, 2015.
\bibitem{Greentouch2017}
Greentouch, “\emph{White Paper on GreenTouch Final Results from Green Meter Research Study,}”  Available from: "https://s3-us-west-2.amazonaws.com/belllabs-microsite-greentouch/uploads/documents/ GreenTouch\_Green\_Meter\_Final\_Results\_18\_June\_2015.pdf"
\end{thebibliography}

%\begin{biography}[{\includegraphics[width=1in,height=1.25in,clip,keepaspectratio]{mshell}}]{Michael Shell}

\begin{IEEEbiographynophoto}{Mohamed Musa}
received the BSc degree (first-class Honours) in Electrical and Electronic Engineering from the University of Khartoum, Sudan, in 2009, the MSc degree (with distinction) in Broadband Wireless and Optical Communication from University of Leeds, UK, in 2011. He received the PhD from University of Leeds in 2016 in energy efficient network coding in optical networks. His current research interests include energy optimization of ICT networks, network coding and energy efficient routing protocols in optical networks. 
\end{IEEEbiographynophoto}

\begin{IEEEbiographynophoto}{Dr. Taisir Elgorashi}
received the B.S. degree (first-class Honors) in electrical and electronic engineering from the University of Khartoum, Khartoum, Sudan, in 2004, the M.Sc. degree (with distinction) in photonic and communication systems from the University of Wales, Swansea, UK, in 2005, and the Ph.D. degree in optical networking from the University of Leeds, Leeds, UK, in 2010. She is currently a Lecturer of optical networks in the School of Electrical and Electronic Engineering, University of Leeds. Previously, she held a Postdoctoral Research post at the University of Leeds (2010–2014), where she focused on the energy efficiency of optical networks investigating the use of renew- able energy in core networks, green IP over WDM networks with data centers, energy efficient physical topology design, energy efficiency of content distribution networks, distributed cloud computing, network virtualization, and big data. In 2012, she was a BT Research Fellow, where she developed an energy efficient hybrid wireless-optical broadband access network and explored the dynamics of TV viewing behavior and program popularity. The energy efficiency techniques developed during her postdoctoral re- search contributed three out of the eight carefully chosen core network energy efficiency improvement measures recommended by the GreenTouch consortium for every operator network world- wide. Her work led to several invited talks at GreenTouch, Bell Labs, the Optical Network Design and Modeling conference, the Optical Fiber Communications Conference, the International Conference on Computer Communications, and the EU Future Internet Assembly in 2013 and collaboration with Alcatel Lucent and Huawei.
 
\end{IEEEbiographynophoto}

%\newpage

\begin{IEEEbiographynophoto}{Prof. Jaafar Elmirghani}
is the Director of the Institute of Integrated Information Systems within the School of Electronic and Electrical Engineering, University of Leeds, UK. He joined Leeds in 2007 and prior to that (2000-2007) as chair in optical communications at the University of Wales Swansea he founded, developed and directed the Institute of Advanced Telecommunications and the Technium Digital (TD), a technology incubator/spin-off hub. He has provided outstanding leadership in a number of large research projects at the IAT and TD. 
He received the BSc in Electrical Engineering, First Class Honours from the University of Khartoum in 1989 and was awarded all 4 prizes in the department for academic distinction. He received the PhD in the synchronization of optical systems and optical receiver design from the University of Huddersfield UK in 1994 and the DSc in Communication Systems and Networks from University of Leeds, UK, in 2014. He has co-authored Photonic Switching Technology: Systems and Networks, (Wiley) and has published over 450 papers. He has research interests in optical systems and networks. 
Prof. Elmirghani is Fellow of the IET, Chartered Engineer, Fellow of the Institute of Physics and Senior Member of IEEE. He was Chairman of IEEE Comsoc Transmission Access and Optical Systems technical committee and was Chairman of IEEE Comsoc Signal Processing and Communications Electronics technical committee, and an editor of IEEE Communications Magazine. He was founding Chair of the Advanced Signal Processing for Communication Symposium which started at IEEE GLOBECOM’99 and has continued since at every ICC and GLOBECOM. Prof. Elmirghani was also founding Chair of the first IEEE ICC/GLOBECOM optical symposium at GLOBECOM’00, the Future Photonic Network Technologies, Architectures and Protocols Symposium. He chaired this Symposium, which continues to date under different names. He was the founding chair of the first Green Track at ICC/GLOBECOM at GLOBECOM 2011, and is Chair of the IEEE Green ICT initiative within the IEEE Technical Activities Board (TAB) Future Directions Committee (FDC), a pan IEEE Societies initiative responsible for Green ICT activities across IEEE, 2012-present. He is and has been on the technical program committee of 34 IEEE ICC/GLOBECOM conferences between 1995 and 2016 including 15 times as Symposium Chair. He has given over 55 invited and keynote talks over the past 8 years.
He received the IEEE Communications Society Hal Sobol award, the IEEE Comsoc Chapter Achievement award for excellence in chapter activities (both in international competition in 2005), the University of Wales Swansea Outstanding Research Achievement Award, 2006; and received in international competition: the IEEE Communications Society Signal Processing and Communication Electronics outstanding service award, 2009, a best paper award at IEEE ICC’2013. Related to Green Communications he received (i) the IEEE Comsoc Transmission Access and Optical Systems outstanding Service award 2015 in recognition of “Leadership and Contributions to the Area of Green Communications”, (ii) the GreenTouch 1000x award in 2015 for “pioneering research contributions to the field of energy efficiency in telecommunications”, (iii) the IET 2016 Premium Award for best paper in IET Optoelectronics and (iv) shared the 2016 Edison Award in the collective disruption category with a team of 6 from GreenTouch for their joint work on the GreenMeter.
He is currently an editor of: IET Optoelectronics and Journal of Optical Communications, an was editor of IEEE Communications Surveys and Tutorials and IEEE Journal on Selected Areas in Communications series on Green Communications and Networking. He was Co-Chair of the GreenTouch Wired, Core and Access Networks Working Group, an adviser to the Commonwealth Scholarship Commission, member of the Royal Society International Joint Projects Panel and member of the Engineering and Physical Sciences Research Council (EPSRC) College.  He has been awarded in excess of £22 million in grants to date from EPSRC, the EU and industry and has held prestigious fellowships funded by the Royal Society and by BT. He is an IEEE Comsoc Distinguished Lecturer 2013-2016.
 
\end{IEEEbiographynophoto}


%\vfill



\end{document}